\documentclass[conference]{IEEEtran}
\IEEEoverridecommandlockouts
\usepackage{subcaption}
\usepackage{caption}
\usepackage{amsmath,epsfig}
\usepackage{bm}
\usepackage{slashbox}
\usepackage{algorithmic}
\usepackage{latexsym}
\usepackage{amsmath}
\usepackage{amssymb}
\usepackage{epsfig}
\usepackage{graphicx}
\usepackage[ruled,vlined]{algorithm2e}
\usepackage{epstopdf}

\begin{document}
\vspace{-8cm}\title{\huge Exploring Social Ties for Enhanced Device-to-Device Communications in Wireless Networks}\vspace{-8cm}

\vspace{-8cm}\author{Yanru Zhang, \thanks{Y. Zhang (yzhang82@uh.edu) and Z. Han (zhan2@uh.edu) are with the Department of Electrical and Computer Engineering, University of Houston, Houston, Texas 77004.}
Lingyang Song,\thanks{L. Song (lingyang.song@pku.edu.cn) is with the School of Electrical Engineering and Computer Science, Peking University, Beijing, China, 100871.}
Walid Saad,\thanks{W. Saad (walid@miami.edu) is with the Electrical and Computer Engineering Department, University of Miami, Coral Gables, Florida 33146.}
Zaher Dawy,\thanks{Z. Dawy (zg03@aub.edu.lb) is with the Electrical and Computer Engineering Department, American University of Beirut, Beirut, Lebanon.}
and Zhu Han
\thanks{This work was made possible by NPRP grant $\sharp4-347-2-127$ from the Qatar National Research Fund (a member of Qatar Foundation). The statements made herein are solely the responsibility of the authors.}} \vspace{-8cm}\medskip\vspace{-8cm}
\vspace{-8cm}\maketitle\vspace{-8cm}

\vspace{-8cm}
\begin{abstract}
Device-to-device (D2D) communications is seen as a major technology to overcome the imminent wireless capacity crunch and to enable novel application services. In this paper, we propose a novel, social-aware approach for optimizing D2D communications by exploiting two network layers: the social network and the physical, wireless network. First we formulate the physical layer D2D network according to users' encounter histories. Subsequently, we propose a novel approach, based on the so-called Indian Buffet Process, so as to model the distribution of contents in users' online social networks. Given the online and offline social relations collected by the Evolved Node B, we jointly optimize the traffic offload process in D2D communication. Simulation results show that the proposed approach offload the traffic of Evolved Node B successfully.
\end{abstract}

\vspace{-0.05cm}
\section{Introduction}
\vspace{-0.15cm}
The recent proliferation of smartphones and tablets is seen as a key enabler for anywhere, anytime wireless communications. The rise of websites, such as Facebook and YouTube, significantly increases the frequency of users' online activities. Due to this continuously increasing demand for wireless access, a tremendous amount of data is circulating over today's wireless networks. This increase in demand is straining current cellular systems, thus requiring novel approaches for network design.

In order to cope with this wireless capacity crunch, device-to-device (D2D) communications underlaid on cellular systems has recently emerged as a promising technique that can significantly boost the performance of wireless networks. In D2D communication, user equipments (UEs) transmit data signals to each other over a direct link instead of through the wireless infrastructure, i.e., the cellular network's Evolved Node Bs (eNBs). The key idea is to allow direct D2D communications over the licensed band and under the control of the cellular system's operator \cite{Xu.2012}. Furthermore, as D2D communication often occurs over shorter distances, it is expected to yield higher data rates for the UEs than infrastructure-based communications. D2D communication is regarded as a promising technology for improving the spectral utilization of wireless systems.

Recent studies have shown that the majority of the traffic in cellular systems consists of the download of content such as videos or mobile applications. If we can cut off the traffic of this part, a large amount of capacity can be freed out. Usually, popular contents, such as certain YouTube videos, are requested by much more frequently than others. As a result, the eNBs often end up serving different mobile users with the same contents for multiple times. As the eNB has already sent the contents to mobile users, the contents are now locally accessible to other users in the same area, if cellular UEs' resource blocks (RBs) can be shared with others. Upcoming users who are within the transmission distance can request the contents from those users through D2D communication. In this case, the eNB will have to only serve those users who request ``new" content, which has never been downloaded before. Through this D2D communication, we can reduce considerable redundant requests to the eNB, so that the traffic burden of eNB can be released.

The main contribution of this paper is to propose a novel approach to D2D communications that allows to exploit the social network characteristics so as to improve the performance and reduce the load on the wireless cellular system. To achieve this goal, there are some key points we are going to discuss. The first of all is to establish a stable D2D subnetwork to maintain the data transmission successfully. A stable connection should guarantee that once the D2D communication is set up, the link should not be dropped easily. As a D2D subnetwork is formulated by individual users, the connectivity among users sometimes is intermittent. If the connection is too sensitive to users' movement and easy to interrupt, it can neither offload the traffic of eNB nor meet users' satisfaction. It is difficult to employ such dynamic information to make reasonable decisions. However, the social relations in real world tend to be stable over time. Such social ties can be utilized to achieve efficient data transmission in the D2D subnetwork. We name this social relation assisted data transmission network by offline social network (OffSN).

Second, we assess the amount of traffic that can be offloaded to D2D communication, i.e., what probability the requested contents can be served locally. To analyze this problem, we study the probability that a certain content is selected. This probability is affected by two different aspects: the external (external influence from media, friends, etc.), and internal (user's own interests) aspects. While users' interests are hard to know, the external influence is more easier to estimate. The choices of the users are mutually dependent. Consequently, the network operator (e.g., via the eNB) can generate an online social network (OnSN) to keep track of users' access to online websites, and maintain the distribution over OffSN users' selection. If we can estimate people's selection based on external influences, we can get the probability of contents being served locally.

In this paper, We will adopt practical metrics to establish OffSN, and involve the novel learning process-Indian Buffet Process to model the influence in OnSN. Then we can get solutions for the previous two problems. Latter we will integrate OffSN and OnSN to carry out the traffic offloading algorithm for the cellular network. Our simulations also proved our analysis for the traffic offloading performance.
\vspace{-0.25cm}
\section{System Model}
\vspace{-0.15cm}
Consider a cellular network with one eNB and $N$ users. The UEs can receive signals from the eNB through cellular network, or from other UEs through D2D pairs using licensed spectrum resources. In the system, two network layers exist over which information is disseminated. The first layer is OnSN. The links of contents spread out on popular websites, users access the links to contents. Hence, the OnSN is the platform over which users acquire the links of contents. Once a link is accessed, the data package of contents must be transmitted to the UEs through the actual physical network. Taking advantage of the social ties, OffSN is the physical layer network for contents behind the links spread out.
\begin{figure}[!t]
  \begin{center}
    \includegraphics[width=0.9\columnwidth,height=0.28\textwidth]{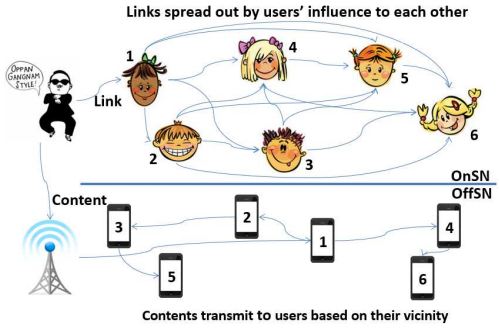}
    \vspace{-0.15cm}
    \caption{\label{fig:OnSN and OffSN}Information dissemination in both OnSN and OffSN.}
  \end{center}\vspace{-0.8cm}
\end{figure}

An illustration of this proposed model is shown in Fig. \ref{fig:OnSN and OffSN}. Each user active in the OnSN corresponds to an UE in the OffSN. Users access the link of a content in an increasing order of their labels. In OnSN, the link of a content is spread out according their popularity from frequent users to regular users. In particular, a group of users, which we refer to as $\emph{frequent users}$, have a high online activity, and, thus, are the main source of influence and information dissemination. In this respect, the choices of the $\emph{regular users}$, who access the OnSN less frequently, are usually influenced by the frequent users. In the OffSN, the first request of the content is served by the eNB. Subsequent coming users can thus be served by previous users who hold the content, if they are within the D2D communication distance. In this section, we will study the properties of OffSN and OnSN, and quantify the relation between these two networks, before developing the proposed offloading algorithm.
\vspace{-0.05cm}
\subsection{Offline Social Network Model}
\vspace{-0.05cm}
In the area covered by an eNB, the distribution of the users can often be properly modeled. For instance, in public areas such as office buildings and commercial sites, the density of users is much higher than in other locations such as sideways and open fields. In addition, users are less likely to browse the web when they are walking or driving. Indeed, the majority of the data transmissions occurs in those fixed places. In such high density locations, forming D2D network as OffSN becomes a natural process. Thus, we can distinguish two types of areas: highly dense areas such as office buildings, and ``white" areas such as open fields. In the former, we assume that D2D networks are formed based on users' social relations. While in the latter, due to the low density, the users are served directly by the eNB. D2D communications carry out in those subnetworks can effectively improve the cellular system throughput. If proper resource management is adopted, the interference among each subnetwork can be restricted \cite{Golrezaei.ICC2012}. Thus, each OffSN can be mutually independent from others to some extent. Optimizing the performance of each OffSN can improve the overall performance of the cellular network.

The OffSN is a reflection of local users' social ties. Proper metrics need to be adopted to depict the degree of the connections among users in the OffSN. For example, users who are within the D2D communication range and more regularly meet can be seen as more robust connection. Due to mobility, grouping users by using only their last known location, can lead to dropped connections. Indeed, in public areas such as airports and train stations, most of the users have high mobility patterns and it is difficult to predict their future locations. However, if we define users connection degree in OffSN according to their encounter history or daily routes, stable D2D connections can be formed. For example, officemates, classmates and family members always meet one another more regularly and frequently than others. Thus, those social ties lead to higher probabilities to transmit data among users. If any two UEs are within the D2D communication distance, the eNB can detect and mark them as encountered. But the specific way to collect data is not our main focus in this paper.

The contact duration distribution between two users is assumed to be a continuous distribution, which has a positive value for all real values greater than zero. For users who regularly meet, their encounter duration usually centers around a mean value. So we can adopt a $Gamma(k,\theta)$ distribution to model the encounter duration between two users. This distribution is widely used in modeling the call durations and has been shown to have a high accuracy \cite{Guo.WiCom 2007}. To find the value for the two parameter $k$ and $\theta$, we need to derive the mean and variance of the contact duration.

\begin{figure}[!t]
  \begin{center}
    \includegraphics[width=0.7\columnwidth,height=0.13\textwidth]{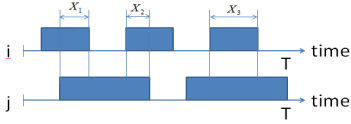}
    \vspace{-0.3cm}
    \caption{\label{fig:encounter history}Encounter history between user i and j.}
  \end{center}\vspace{-0.7cm}
\end{figure}

As shown in Fig. \ref{fig:encounter history}, given the contact duration $X_n$ and the number of encounters $N_{i,j}$ between user $i$ and user $j$, an estimate of the expected contact duration length $M_{i,j}$:
\vspace{-0.2cm}
\begin{equation}\vspace{-0.2cm}
M_{i,j} = \frac {\sum_n X_n}{N_{i,j}}.
\end{equation}\vspace{-0.2cm}

The variance represents the fluctuation in the contact period. If two cases have the same average contact period, the one with larger fluctuations would be less preferable since the contact length would be more uncertain. Thus, we measure the variance of the contact period distribution to reflect the fluctuation using an irregularity metric $I_{i,j}$ defined by \cite{Li.IEEE Conf2009}:
\vspace{-0.2cm}
\begin{equation}\vspace{-0.25cm}
I_{i,j} = \frac {\sum_n (X_n - M_{i,j})^2} {N_{i,j}}.
\end{equation}\vspace{-0.15cm}

Given the mean and variance of the encounter period, we can derive the encounter duration distribution: $X\sim Gamma(k,\theta)= Gamma(\frac{M_{i,j}^2}{I_{i,j}}, \frac{I_{i,j}}{M_{i,j}})$. The probability density distribution (PDF) of encounter duration is:
\vspace{-0.1cm}
\begin{equation}\vspace{-0.2cm}
f(x; k, \theta) = \frac{1}{\theta^k}\frac{1}{\Gamma(k)}x^{k-1}e^{-\frac{x}{\theta}},
\end{equation}\vspace{-0.05cm}
where $\Gamma(k)=\int_0^\infty t^{k-1}e^{-t}dt$. Then, we can calculate the probability of the contact durations that are qualified for data transmission. If the contact duration is not enough to complete a data package transmission, the communication session cannot be carried out successfully. We adopt a closeness metric, $w_{i,j}$ to represent the probability of establishing a successful communication period between two UEs $i$ and $j$, which ranges from $0$ to $1$. The qualified contact duration is the complementary of the disqualified communication duration probability, so $w_{i,j}$ can be represented as:
\vspace{-0.1cm}
\begin{equation}\vspace{-0.25cm}
w_{i,j}=1-\int_0^{X_{min}}f(u;k,\theta)du=1-\frac{\gamma(k,\frac{X_{min}}{\theta})}{\Gamma(k)},
\end{equation}\vspace{-0.05cm}
where $X_{min}$ is the minimal contact duration required to successfully transmit one content data package. $X_{min}$ is a random variable that depends on the channel conditions between the two UEs (e.g., the higher the signal strength between the UEs, the smaller $X_{min}$). Moreover, $X_{min}$ should depend on the content size. $\gamma(k,\frac{X_{min}}{\theta})=\int_0^{X_{min}} t^{k-1}e^{-t}dt$ is the lower incomplete gamma function. Larger closeness $w_{i,j}$ indicates a better future contact opportunity between user $i$ and user $j$.

An OffSN can be seen as a D2D network that is constructed by a group of users with stable connections. Hence, we can use the closeness metric $w_{i,j}$ to describe the communication probability between two users, which can also be seen as the weight of the link between user $i$ and user $j$. Then, a threshold $w_T$ can be defined to filter the boundary between different OffSNs and ``white" areas.

If we have lower $w_T$, more users will be added in, and the covered area of the OffSN will increase. Even though the probability that a user will be served locally will increase, it is likely that the group members who own the content are located far away. In this case, the prospective gain would not compensate the cost on associated power consumption for relaying and transmitting. Also, OffSNs may overlap with each other, and cause inter-OffSN interference which can further complicate the analysis. On the other hand, if $w_T$ is too high, only a small number of users can be grouped which makes it difficult to perform D2D communication. Therefore, one important problem is to find a proper $w_T$ that can balance the tradeoff between the cost and gain, and thus to get the best performance for the system.
\vspace{-0.05cm}
\subsection{Online Social Network Model}
\vspace{-0.05cm}
First, we define the number of users in OnSN as $N$ which is corresponding to $N$ UEs in OffSN. The total number of available content in OnSN is denoted by $K$. Given the large volume of data available online, we can assume that $K = K_h + K_0$, $K \rightarrow \infty$. $K_h$ represents the set of contents that have viewing histories and $K_0$ is the set of contents that do not have any viewing history. In OnSN, users select contents partially based on external influence. It is possible to predict current users' selections by analysing the OnSN activities of previous users within the same OffSN. To draw the probability of current user's selection, we adopt the Indian Buffet Process (IBP) \cite{Griffiths.ML 2011} which serves as a powerful analytical tool for predicting users' behaviors.

In an IBP, there are infinite dishes for customers to choose. The first customer will select its preferred dishes according to a $Poisson$ distribution with parameter $\alpha$. Since all dishes are new to this customer, no reference or external information exists so as to influence this customer's selection. However, once the first customer completed the selection, the following customers will have some prior information about those dishes based on the first customer'’s feedback. Therefore, the decisions of subsequent customers are influenced by the previous customers'’ feedback. Customers learn from the previous selections to update their beliefs on the dishes and the probabilities with which they will choose the dishes.

The behavior of content selection in OnSN is analogous to the behavior of dish selection in an IBP. If we view OnSN as an Indian buffet, the online content as the infinite number of dishes, and the users as customers, we can interpret the contents spreading process online by an IBP. Users enter OnSN sequentially to download their desired content. When a user downloads its content, the recorded downloading times of contents changed. This action will affect the probability that this content to be requested. Popular contents will be requested more frequently. While those contents that are only favored by a few number of people, or those new produced content will be requested less frequently. So the probability distribution can be implemented from the IBP directly.

\begin{figure}[!t]
  \begin{center}
    \includegraphics[width=\columnwidth]{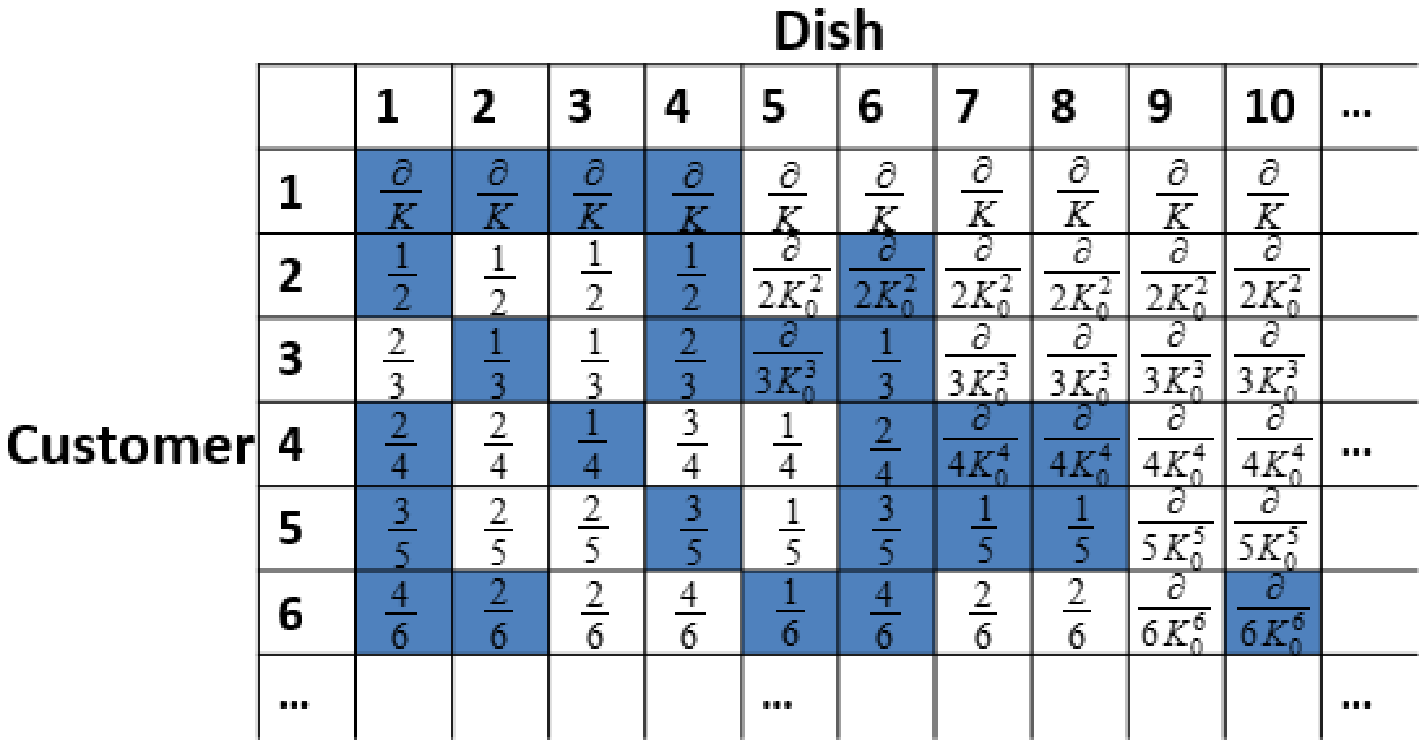}
    \vspace{-0.3cm}
    \caption{\label{fig:IBP}Indian Buffet Process.}
  \end{center}\vspace{-0.7cm}
\end{figure}
In Fig. \ref{fig:IBP}, we show one realization of IBP. Customers are labeled by ascending numbers in a sequence. The shaded block represent the $n$th user selected dish $k$. In IBP, the first customer selects each dishes with equal probability of $\frac{\alpha}{K}$, and ends up with the number of dishes follows $Poisson(\alpha)$ distribution. For subsequent customers $n = 2,\ldots,N$, the probability of also having dish $k$ already belonging to previous customers is$ \frac{m_k^{n-1}}{n}$, where $m_k^{n-1}$ is the number of customers prior to $n$ with dish $k$ \cite{Griffiths.ML 2011}. Repeating the same argument as the first customer, customer $n$ will also have $m_n^0$ new dishes not tasted by the previous customers follows a $Poisson(\frac{\alpha}{n})$ distribution.

The probabilities of selecting certain dishes act as the prior information $\pi(m_k^{n-1})$. For ``old" dishes which have been tasted before, $m_k^{n-1}\neq 0$. For ``new"’ dishes which have not been sampled before, $m_k^{n-1}= 0$. After user $n$ completes its selection, the prior will be updated to $\pi(m_k^{n})$. This learning process is also illustrated in Fig. \ref{fig:IBP}. $K_0^n$ is the number of dishes that have not been sampled before user $n$'s selection session. We can see the selection probability for dish $k$ updated every time after each customer's selection.
\vspace{-0.03cm}
\section{Proposed Traffic Offloading Algorithm}
\vspace{-0.03cm}
\subsection{Data Rate in OffSN}
\vspace{-0.05cm}
As the inter-OffSN interference of D2D communication can be restricted by methods such as power control \cite{Xu.2012}\cite{Golrezaei.ICC2012}. For simplicity, we place an emphasis on the intra-OffSN interference due to resource sharing between D2D and cellular communication. In the OffSN that acts as the subnetwork of cellular network, the D2D transmissions and the eNB transmissions will interfere. During the downlink period of D2D communication, UEs will experience interference from other cellular and D2D communications as they share the same subchannels. In this respect, the received power of link $i$ to $j$ can be expressed as \cite{Xu.2012}:
\begin{equation}\vspace{-0.25cm}
P_{r,ij}= P_i\cdot h^2_{ij}=P_i\cdot d_{ij}^{-\eta}\cdot h^2_0,
\end{equation}\vspace{-0.05cm}
where $P_i$ is the transmit power, $\eta$ is path loss exponent which ranges from $2 \leq \eta \leq 5$, $h_{ij}$ is the channel response of
the link $i$ to $j$, $h_0$ is the complex Gaussian channel coefficient that follows the complex normal distribution $\textsl{CN}(0,1)$.

Thus, we can define the transmission rate of users served by the eNB and by D2D communication with co-channel interference \cite{Xu.2012}:
\begin{equation}\vspace{-0.25cm}
R_c= \log_2 \left(1+\frac {P_Bh_{Bc}^2}{\sum_d\beta_{cd}P_dh_{dc}^2+N_0}\right),
\end{equation}
\begin{equation}
R_d= \log_2 \left(1+\frac {P_dh_{dd}^2}{P_Bh_{Bd}^2+\sum_{d'}\beta_{dd'}P_{d'}h_{d'd}^2+N_0}\right),
\end{equation}
where $P_B$, $P_d$ and $P_{d'}$ are the transmit power of eNB, D2D transmitter $d$ and $d'$, respectively, $N_0$ is the additive white Gaussian noise (AWGN) at the receivers, and $\beta_{cd}$ represents the presence of interference from D2D to cellular communication, satisfying $\beta_{cd} = 1$, otherwise $\beta_{cd} = 0$. Here, $d\neq d'$, so $\sum_{d'}\beta_{dd'}P_{d'}h_{d'd}^2$ represents the interference from the other D2D pairs that share spectrum resources with pair $d$.

The transmission rate of the users that are only serviced by the eNB and that do not experience co-channel interference in the OffSN is given by:
\begin{equation}\vspace{-0.25cm}
V_c= \log_2 \left(1+\frac {P_Bh_{Bc}^2}{N_0}\right).
\end{equation}\vspace{-0.1cm}
\vspace{-0.05cm}
\subsection{Content Selection in OnSN}
\vspace{-0.05cm}
In the studied model, the eNB maintains the contents distribution of each OnSN. When the $n$th user starts to surf online, this user will sample the content based on the prior information $\pi(m_k^{n-1})$. To this end, this user will access old content with probability $\frac{m_k^{n-1}}{n}$, and access new content with a $Poisson(\frac{\alpha}{n})$ distribution. After content selection, $\pi(m_k^{n-1})$ is updated to the posterior probability $\pi(m_k^n)$. The total amount of content each user selected can be draw from a $Poisson$ distribution
\begin{equation}\vspace{-0.25cm}
m_n \sim Poisson(\alpha).
\end{equation}\vspace{-0.3cm}
\vspace{-0.15cm}
\subsection{System Utility}
\vspace{-0.1cm}
In the proposed model, the users aim at maximizing their data rate while minimizing the cost. Without loss of generality, we assume that every content has the same size. Consequently, we propose the following utility function for user $n$:
\begin{equation}\vspace{-0.25cm}
U_u(n)=\left[\sum_{k=1}^{K_h}\frac {m_k^{n-1}}{n}\right] (R_d-C_t)+m_n^0 (R_c- C_m),
\end{equation}
where $C_t$ is the cost related to the D2D transmission power $P_i$ in (6). $C_m$ is the cost paid by the user  to the eNB for the data flow. The utility function consists of two parts. The first part is the utility of receiving old content via D2D communication. The second part is the utility of downloading new content from the eNB.

From the eNB perspective, the goal is to maximize the overall data rate while also offloading as much traffic as possible. Even though eNB can offload traffic by D2D communication, controlling the switching over cellular and D2D communication causes extra data transmission. There thus exists some cost such as control signals transmission and information feedback during the access process \cite{Xu.2012}. Therefore, for the eNB that is servicing a certain user $n$, we propose the following utility function:
\begin{equation}\vspace{-0.15cm}
U_B(n)=\left[\sum_{k=1}^{K_h}\frac {m_k^{n-1}}{n} \right]R_d+m_n^0 R_c - m_n C_c,
\end{equation}
where $C_c$ is the cost for controlling the resource allocation process. The total traffic eNB offloaded by D2D communication is $m_n V_c-m_n C_c-m_n^0 R_c$.
\vspace{-0.2cm}
\subsection{Proposed Algorithm}
\vspace{-0.5cm}
\begin{algorithm}
\caption {Proposed Traffic Offloading Algorithm}
\SetAlgoLined
\KwData {$X_n$, $N_{i,j}$, $P_B$, $P_d$, $P_{d'}$}
\KwResult{$U_u(n)$, $U_B(n)$}
\textbf{1. OffSN Generation}\;
eNB collects encounter information in cellular network\;
Locate the frequent users in high user density areas\;
Find the closeness $w_{i,j}$ between two users\;
\If {$w_{i,j}\geq w_T$}{
Add user $i$ and user $j$ into OffSN\;
}
\textbf{2. User Activity Detection}\;
\While {user $n$ is accessing the tagged websites}{
eNB detects user's location\;
 \If {user $\in$ ``white" areas}{
  eNB serves all requests directly\;
 }
 \If{user $\in$ an OffSN}{
 eNB keep watching user's activities\;
 }
}
\textbf{3. Service based on OnSN Activities}\;
\While{User $n$ is selecting contents online}{
eNB maintains prior information $\pi(m_k^{n-1})$\;
\If {old content}{
eNB locates the content holder with highest closeness $w_{n,i}$ to user $n$ in OffSN\;
Establish D2D communication\;
eNB monitors the communication process\;
 \If {communication failed}{
 eNB revert back to continue the transmission\;
 }
}
\If {new content}{
 eNB serves the request directly\;
 }
update $\pi(m_k^{n})$\;
}
\vspace{-0.15cm}\end{algorithm}
\vspace{-0.35cm}
We propose a novel algorithm form which consists of multiple stages. In the first stage, the eNB focuses on high user density areas, and collects the encounter history between users. The eNB locates one frequent user and its neighboring users via well-known algorithms \cite{{Irfan.2013}} in order to compute their closeness $w_{i,j}$. By checking if $w_{i,j}\geq w_T$, i.e., if user $i$ and user $j$ satisfy the predefined closeness threshold, the eNB can decide on whether to add this user into the OffSN or not. By choosing a proper $w_T$ and power control, the interference among different OffSNs can be avoided. This process will continue until no more user can be further added to the eNB's list. Then, the users in the established OffSN can construct a communication session with only intra-OffSN interference.

For websites that provide a portal to access content, such as Facebook and YouTube, the eNB will assign a special tag. Once a user visits such tagged websites, the eNB will inspect whether the user is located in an OffSN or a ``white" area. If the user is in ``white" area, user's any requests will be served by the eNB directly. If the user is located in an OffSN, the eNB will wait until the user requests contents. By serving previous user's requests, the eNB has already built up a history file including the prior information $\pi(m_k^{n-1})$ of the content distribution in the OnSN. As soon as receiving user requests data, the eNB detects if there are any resources in the OffSN, and then choose to set up a D2D communication or not based on the feedback. For old content, the eNB will send control signal to the $i$th UE with the highest closeness $w_{n,i}$ with user $n$. Then UE $n$ and UE $i$ establish the D2D communication. Even if the D2D communication is setup successfully, the eNB still wait until the process finishes successfully. If the D2D communication fails, the eNB will revert back to serving the user directly. For new content, the eNB serves the user directly. After the selection is complete, the prior information updates to the posterior probability $\pi(m_k^n)$. The proposed D2D communication algorithm is summarized in Algorithm $1$.
\vspace{-0.15cm}
\section{Simulation Results and Analysis}
\vspace{-0.1cm}
In this section, we give the simulation results to show traffic offloading performances of the system from different aspects and provide numerical results of how different parameters affect the system's performance. Consider $32$ active users than are randomly distributed in an OffSN. The size of the content library is unbounded. We assume that the content selection process has been carried on for a period of times. So the eNB can obtain the prior information of the content distribution. In our simulation, we have proved that, once the parameters are specified, the order in which the users perform their selection does not affect the performance of the system.
The main physical layer parameters are listed as follows. The radius of an OffSN is set up as $80$m. Noise spectral density is $-174$dBm/Hz. Noise figure at device is $9$dB. Antenna gains and transmit power of eNB is $14$dBi and $46$dBm. For device, it is $0$dBi and $23$dBm, respectively.

\begin{figure}[!t]
  \begin{center}
    \includegraphics[width=0.825\columnwidth,height=0.275\textwidth]{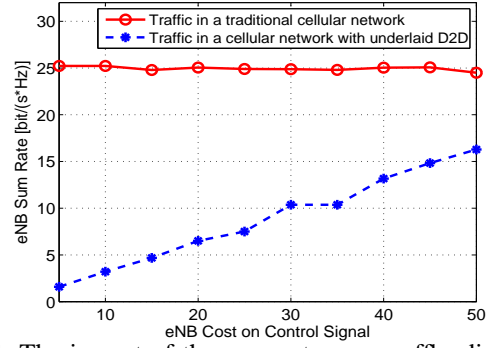}
    \vspace{-0.25cm}
    \caption{\label{fig:Traffic and Alpha}The impact of the parameter $\alpha$ on offloading traffic.}
  \end{center}\vspace{-0.5cm}
\end{figure}

In Fig. \ref{fig:Traffic and Alpha}, we investigate whether the user online activity degree will affect the amount of traffic that can be offloaded. As we can see from Fig. \ref{fig:Traffic and Alpha}, with the increase of parameter $\alpha$, the amount of traffic that is offloaded from the eNB decreases, but traffic still can be released compared to the eNB serving only system. The data rate is nearly $0$ when $\alpha$ is small. This result is due to the fact that, when $\alpha$ is low, users are more likely to choose old content. Here, the requests are served by D2D communication in most of the cases, and cause nearly no traffic on the eNB. As $\alpha$ increases, the data rate begins increase. Indeed, when the users tend to make more selections, they may choose more new content. The offloaded traffic amount generally decreases with the increase of online activities, which is coincidence with the common sense that, more contents downloading will cause more traffic to eNB. The traffic includes not only the contents data, but also the control signals the eNB needs to send for the D2D communication arrangement.
\begin{figure}[!t]
  \begin{center}
    \includegraphics[width=0.825\columnwidth,height=0.275\textwidth]{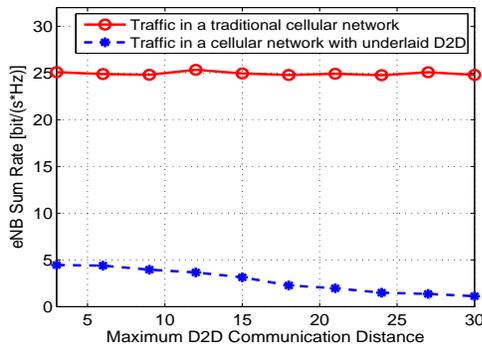}
    \vspace{-0.25cm}
    \caption{\label{fig:Traffic and Distance}The relationship between the offloading traffic and maximum D2D communication distance.}
  \end{center}\vspace{-0.45cm}
\end{figure}

As the D2D communication distance increases, the eNB will have more possibilities for detecting available contents providers. As a result, the performance of traffic offloading will be better with larger maximum distance. This assumption is shown in Fig. \ref{fig:Traffic and Distance}. In this figure, we can see that, with $\alpha$ setting to $8$, increasing the maximum communication distance, yields a decrease in the eNB's data rate and an increase in the amount of offloaded traffic. However, we note that, with the increase of the transmission distance, the associated UE costs (e.g., power consumption) will also increase. Thus, the increase of D2D communication distance will provides additional benefits to the eNB, but not for users.

\begin{figure}[!t]
  \begin{center}
    \includegraphics[width=0.825\columnwidth,height=0.275\textwidth]{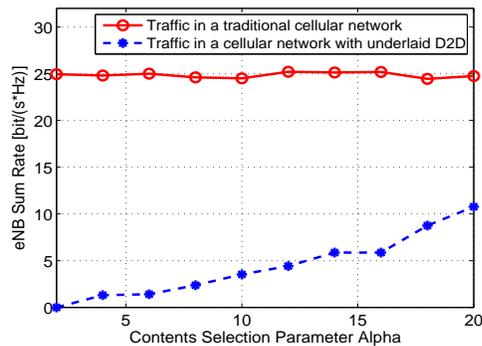}
    \vspace{-0.25cm}
    \caption{\label{fig:Traffic and eNB Cost}Average sum-rate at the eNB, as the cost for control signaling varies.}
  \end{center}\vspace{-0.85cm}
\end{figure}
In Fig. \ref{fig:Traffic and eNB Cost}, we show the variation of the sum-rate at the eNB as the cost for control signal varies. As the eNB has to arrange the inter change process between cellular and D2D communication, necessary control information are needed. Moreover, additional feedback signals are required for monitoring the D2D communication and checking its status. Those costs will affect the traffic offloading performance of the system. In our simulation, we define the cost as the counteract to the gain in data rate from 5\% to 50\%.  As we can see from Fig. \ref{fig:Traffic and eNB Cost}, increasing the cost on control signal, the offload traffic amount is decreased.
\vspace{-0.05cm}
\section{Conclusion}
\vspace{-0.05cm}
In this paper, we have proposed a novel approach for improving the performance of D2D communication underlaid over a cellular system, by exploiting the social ties and influence among individuals. We formed the OffSN to divide the cellular network into several subnetworks for carrying out D2D communication with only intra-OffSN interference. Also we established the OnSN to analyse the OffSN users' online activities. By modeling the influence among users on contents selection online by Indian Buffet Process, we obtain the distribution of contents requests, and thus can get the probabilities of each contents to be requested. With the algorithm we proposed, the traffic of eNB has been released. Simulation results have shown that different parameters for eNB and users will lead to different traffic offloading performances. Users with more online activities will increase and fluctuate the data rate of the eNB. Enable larger maximum D2D communication distance will release more traffic burden of the eNB. While if the cost on D2D communication arrangement is high, the traffic offloading performance will decrease.

%


\end{document}